\documentclass{amsart}

\usepackage{amscd}
\usepackage{bbm}
\usepackage{graphicx}
\usepackage{verbatim}

\usepackage[latin1]{inputenc}

\newtheorem{prop}{Proposition}

\newtheorem{defi}[prop]{Definition}
\newtheorem{theo}[prop]{Theorem}

\def\P{{\mathbb P}}
\def\R{{\mathbb R}}
\def\E{{\mathbb E}}
\def\C{{\mathbb C}}
\def\N{{\mathbb N}}
\def\Z{{\mathbb Z}}

\def\cal{\mathcal}

\def\etal{{\em et al.\ }}
\def\ind{\mathbbm{1}}

\newcommand\Pois[2]{\frac{#1^{#2}}{#2!}\,e^{-#1}}

\title{On the asymptotic behavior of some Algorithms}

\author{Philippe Robert}

\address[Ph.~Robert]{INRIA, domaine de Voluceau, B.P.~105, 78153 Le Chesnay Cedex, France}
\email{Philippe.Robert@inria.fr}
\urladdr{http://www-rocq.inria.fr/\~{}robert}

\date{\today}


\begin{document}
\begin{abstract}
A simple approach is presented to study the asymptotic behavior of some algorithms with an
underlying tree structure.  It is shown  that some asymptotic oscillating behaviors can be
precisely analyzed without  resorting to complex analysis  techniques as it is
usually done in this context. A new explicit representation of periodic functions involved
is obtained at the same time.
\end{abstract}

\maketitle

\bigskip

\hrule

\vspace{-1cm}

\tableofcontents

\vspace{-1cm}

\hrule

\bigskip

\section{Introduction}
Algorithms with an underlying tree structure are quite common in computer science, they
are used to sort, store and search\ldots See Cormen
\etal~\cite{Cormen:01} and Knuth~\cite{Knuth:01}.  Splitting algorithms are 
examples of such algorithms, they can be described  as follows

\bigskip

\noindent
{\sc Splitting Algorithm} ${\cal S}(n)$
\hrule
\begin{itemize}
\item[---] {\sc Termination Condition}.\\
For some subset $F$ of $\N$,  if $n\in F$ \\
\ \phantom{For some subset}  $\longrightarrow$ {\sc Stop}.
\item[---] {\sc Tree Structure}.\\
Randomly divide $n$ into $n_1$,\ldots, $n_d$, with $n_1+\cdots+n_d=n$.\\
\ \phantom{For some subset} $\longrightarrow$ {\sc Apply} ${\cal S}(n_1)$, ${\cal S}(n_2)$, \ldots, ${\cal S}(n_d)$.
\end{itemize}

\hrule
\bigskip

To the algorithm ${\cal S}(n)$ is attached a {\em cost} $R_n$ which can be the number of steps
required to terminate for example. This cost is assumed to be {\em additive} that is, with the above
notations, the relation
\begin{equation}\label{Cost}
R_n=1+  R_{n_1}+R_{n_2}+\cdots+R_{n_d},
\end{equation}
holds.  Note  that $R_n$  is  a {\em  random
variable}  since the  algorithm  divides randomly  into  subgroups. This  quantity can  be
thought  as  the cost  of  processing  $n$ items.  If  $\E(R_n)$  is  its expected  value,
$\E(R_n)/n$ is  the average processing  time of one  item among $n$. From  a probabilistic
point of view,  it  is natural to expect that the
sequence $(R_n)$  satisfies a kind of law of  large  numbers, i.e.  that  $(\E(R_n)/n)$
converges  to  some  quantity $\alpha$. The constant $\alpha$ would be, in some sense,
the asymptotic average processing time of an item. In the context of a communication
network, $R_n$ is the total transmission time of $n$ initial messages, $1/\alpha$ would be
the asymptotic throughput of the protocol. 

Curiously, this law of large numbers does not always hold. In some situations, the sequence
$(\E(R_n)/n)$ does not converge at all and, moreover, exhibits an oscillating behavior. Up
to now, these phenomena have been mostly analyzed by using sophisticated complex
analysis techniques (via functional transforms) by Knuth~\cite{Knuth:01}, Flajolet
\etal~\cite{Flajolet:14} and many others. See  Mahmoud~\cite{Mahmoud:02} for a
comprehensive treatment of this approach. For alternative methods using real
analysis on related problems, see the nice paper by Delange~\cite{Delange:01} and also
Pippenger~\cite{Pippenger:03}.   

In this paper, a direct, simple, approach is proposed to study these uncommon laws of large
numbers. At the same time, it sheds a new light on the oscillating phenomena involved.
First, the classical approach, i.e. with complex analysis, is briefly recalled.  

\subsection*{Analytic approach: An excursion inside the complex plane}
A classical method to derive asymptotics of a sequence $(r_n)$  related to a splitting
algorithm consists in taking successive transforms:
\begin{equation}\label{PoisMell}
(r_n)\quad \stackrel{\text{\small Poisson}}{\longrightarrow}\quad r(x)=\sum_{n\geq 0}r_n\Pois{x}{n}
\quad\stackrel{\text{\small Mellin}}{\longrightarrow}\quad
r^*(s)=\int_0^{+\infty} r(x)x^{s-1}\,dx.
\end{equation}
The Poisson transform step may be, sometimes,  skipped:
\begin{equation}\label{MMell}
(r_n) \quad\stackrel{\text{\small Mellin}}{\longrightarrow}\quad
r^*(s)=\sum_{n\geq 1} r_n \frac{1}{n^s}.
\end{equation}
In this way, Equation~\eqref{Cost} can be translated, via a Poisson
transform, into a functional equation  of the form
\begin{equation}\label{Func}
r(x)=r(\beta x)+h(x), \quad x\geq 0,
\end{equation}
where $0<\beta<1$ and $h$ is some fixed function. Provided that an iteration scheme is valid,
the problem is then to get an asymptotic expansion of the series
\[
r(x)=r(0)+\sum_{n\geq 0} h(\beta^n x)
\]
as $x$ goes to infinity. This is usually done by taking the Mellin transform of the
function $x\to r(x)$.

\subsubsection*{Mellin Transform}
The Mellin transform $r^*$ is defined  in a vertical strip of $\C$ and, under some
growth conditions,  the asymptotic behavior  of $r_n$ [resp.  $r(x)$], as $n$  [resp. $x$]
gets large, can  be expressed by using the poles  of $r^*$ on the right  of the strip
(provided that some growth conditions of the Mellin transform are satisfied). See
Flajolet \etal~\cite{Flajolet:14}.  When  the Poisson transform is used,  the next step is
to justify  that the behavior of  $r(x)$ as $x$ gets  large is similar to  the behavior of
$r_n$ for $n$ large.

\subsubsection*{Poisson Transform}
If  $(N([0,u]), u\geq 0)$ is a Poisson process with parameter $1$ (see the
Appendix for a quick presentation): for $u\geq 0$,  the variable
$N([0,x])$  has a Poisson distribution with parameter $x$.  The function $x\to r(x)$ can
therefore be expressed as  
\[
r(x)=\E\left(r_{N([0,x])}\right),
\]
for  $x\geq 0$. The law of large numbers for Poisson processes states that, almost surely,
$N([0,x])/x\sim 1$ as $x$ gets large, this suggests that $r_{N([0,x])}\sim r_{\lfloor
  x\rfloor}$ provided that the sequence $(r_n)$ does not vary too much: Recall that, due to
the central limit theorem, the approximation $N([0,x])\sim x$ is valid with an error of
the order of $\sqrt{x}$. Analytically, some conditions on the sequence $(r_n)$ can be
formulated so that an equivalence between the asymptotic behaviors of the sequence and of
the Poisson transform can be established. See Jacquet and Szpankowski~\cite{JacSPa} for
example. 

\subsection*{A direct method}
The  approach presented  here relies  on a  convenient use  of Fubini's  Theorem combined,
sometimes, with some elementary properties of  Poisson processes. The purpose of the paper
is to  show that  some asymptotic results for  algorithms on  trees can be  obtained in  an
elementary way.  To  show the effectiveness of the method, expansions with oscillating
behaviors, which are usually  analyzed with quite  technical results of  complex analysis,
are obtained with this approach. 

With  the  analytical approach,  oscillating  expansions  are  described with  a  periodic
function which shows up through its Fourier  coefficients.  It occurs
when the  Mellin transform of the  sequence has poles on  an imaginary axis  at the points
$(a+inb, n\in\Z)$.  The  method presented here has the advantage of  being more direct and
to give an explicit expression of the mysterious periodic function which, in the end, is
not mysterious at all.

Rather than setting up a framework with formal results, the presentation of some important
and   interesting   algorithms   already    analyzed   in   the   literature   has   been
chosen.  Section~\ref{Knuth} studies  one  of the  first  algorithms for  which a  curious
oscillating behavior has  been proved (by Knuth). Section~\ref{Ether}  considers the basic
algorithm of the Ethernet  protocol. It is not, strictly speaking, a  law of large numbers
setting but a similar  oscillating behavior occurs also in this case.  Moreover, it is not
only  true   for  the  averages   but  also  for   the  distribution  of   the  variables.
Section~\ref{Gen} gives other examples where similar methods can be used. In particular, a
treatment of  Equation~\eqref{Func} is proposed.  The examples  of Section~\ref{Ether} and
Section~\ref{Gen} show also that  the method does not give only a  first order term of the
asymptotic expansion  (i.e. at the level  of the law of  large numbers) but  can also give
subsequent terms of the expansion.  A  more complicated splitting algorithm is analyzed in
Mohamed and Robert~\cite{Mohamed:01}.

\subsection*{Acknowledgements}
The author is grateful to Philippe Flajolet for uncountable interesting conversations on tree
algorithms and Mellin transforms. The paper also benefited from two anonymous referees'
comments. 

\section{A Binary Splitting algorithm}\label{Knuth}
This section analyzes an algorithm investigated by Knuth in 1973, see Knuth~\cite{Knuth:01}
page~131-132. It consists in splitting randomly and recursively a group of $n$ initial items
into subgroups until each of the subgroups has cardinality $0$ or $1$. Latter, it has been
used by Capetanakis~\cite{Capetanakis:01} and  Tsybakov and Mikhailov~\cite{Tsybakov:01} in
the design of algorithms for access protocols to communication channels. See also the
surveys Flajolet and Jacquet~\cite{RFlajolet:02} and   Ephremides and
Hajek~\cite{Ephremides:01}. 

\bigskip
\noindent
{\sc Binary Splitting Algorithm} ${\cal K}(n)$

\smallskip
\hrule
\begin{itemize}
\item[---] $n=0$ or $1$ $\longrightarrow$ {\sc Stop}.
\item[---] $n\geq 2$. The $n$ elements are randomly, equally, divided into  two
  subgroups. If  $n_1$ and $n-n_1$ denote the cardinalities of these subgroups then \\
\ \phantom{$n=0$ or $1$} $\longrightarrow$ {\sc Apply} ${\cal K}(n_1)$ and ${\cal K}(n-n_1)$.
\end{itemize}
\hrule

\bigskip 

The cost $R_n$  of the algorithm ${\cal K}(n)$  starting with $n$ elements is  defined as the
number of steps it requires to  stop, in particular, $R_0=R_1=1$. The following proposition
establishes a  recurrence relation  for the  other values of  $n$. It  is simply  a formal
rephrasing of the description of the algorithm.
\begin{prop}[Stochastic Recurrence Relation]\label{CSMA-recprop}
For $n\geq 2$, the random variable $R_n$ satisfies the relation
\[
 R_n\stackrel{\text{dist.}}{=} 1+ R_{n-S_n}^1+R_{S_n}^2, \qquad n\geq 2,
\]
with $S_n=B_1+\cdots+B_n$ and
\begin{itemize}
\item[---] the  i.i.d. variables $B_i$, $i\geq 1$ are Bernoulli with parameter $1/2$;
\item[---] for $0\leq k \leq n$, the variables $R_k^1$ and $R_{n-k}^2$ are independent and
  $R_k^1$ and $R_k^2$ have the same distribution as $R_k$. 
\end{itemize}
\end{prop}
The Bernoulli variables have the following interpretation: For $i\geq 1$, if $B_i=0$
[resp. $B_i=1$ ] the $i$th element goes in the first [resp. second] subgroup. 

The recurrence relation for  $(R_n)$ and the boundary conditions at $n=0$ and $1$ can be
integrated in the following way,
\begin{equation}\label{rec1}
 R_n\stackrel{\text{dist.}}{=} 1+ R_{n-S_n}^1+R_{S_n}^2 -2_{\{n\leq 1\}}.
\end{equation}
The rest of the section is devoted to the analysis of the asymptotic behavior of the
average value $\E(R_n)$ of the random variable $R_n$ when $n$ is large. 
\begin{prop}[Poisson Transform]\label{PoisTransf}
For $x>0$,  the Poisson transform $r(x)$ of the sequence $(\E(R_{n}))$  
is given by the series
\begin{equation}\label{serie}
r(x)=\sum_{n=0}^{+\infty} \E(R_n)\Pois{x}{n}= 1+2\sum_{k\geq 0} 2^{k}\,\P\left(t_2\leq x/2^k\right),
\end{equation}
where $t_2$ is the sum of two independent exponentially distributed random variables with
parameter $1$.
\end{prop}
\begin{proof}
If ${\cal N}=(N([0,t]))$, $(t_n)$ is a Poisson process with intensity $1$. See the appendix.
The Poisson transform $r(x)$ can be expressed as
\[
r(x)=\E\left(R_{N([0,x])}\right),
\]
For $n\geq 0$, $S_n$ denotes the $n$th partial sum of Bernoulli random variables with parameter $1/2$.
Relation~\eqref{rec1} gives the identity
\[
R_{N([0,x])}\stackrel{\text{dist.}}{=}1+R_{S_{N([0,x])}}^1+R_{N([0,x])-S_{N([0,x])}}^2-2_{\{N([0,x])\leq 1\}}.
\]
The  Poisson  variable  random  variable  $N([0,x])$  is split  into  two  random  variables
$S_{N([0,x])}$ and $N([0,x])-S_{N([0,x])}$.  According to Proposition~\ref{PoisSplit} of the
appendix, the  distribution of  these two (independent)  random variables is  Poisson with
parameter $x/2$, i.e. the distribution of $N([0,x/2])$.  Since, by definition, $t_2$ is the second point of the Poisson process
then $\{N([0,x])\leq 1\}=\{t_2>x\}$ and
\[
\E\left(R_{N([0,x])}\right)= 2\E\left(R_{N([0,x/2])}\right)+1 -2\P\left(t_2>x\right).
\]
If $f(x)=r(x)-1$,  the last identity can be written as $f(x)= 2f(x/2)+2\P\left(t_2\leq x\right)$. 
By iterating, one gets the expansion
\[
f(x)=2^nf(x/2^n)+ 2\sum_{k=0}^{n-1}2^k\,\P(t_2\leq x/2^k).
\]
The limit of the sequence $(2^nf(x/2^n)/x)$ is the limit of $(r(h)-1)/h$ as $h$
goes to $0$:
\[
\lim_{n\to+\infty} 2^nf(x/2^n)/x= r'(0)=\E(R_0)-\E(R_1)=0.
\]
Representation~\eqref{serie} of the Poisson transform is thus established.
\end{proof}

The following proposition establishes a useful integral representation of the quantity
$\E(R_n)$ which gives the key of the asymptotic expansion. Its proof uses a probabilistic
trick of independent interest to invert the Poisson transform. 
\begin{prop}[Probabilistic de-Poissonnization]\label{dePois}
For $n\geq 2$,  
\begin{equation}\label{recmoy2}
\E(R_{n}) =4n\int_0^1 2^{-\left\{{-\log_2(x)}\right\}}(n-1)\left(1-x\right)^{n-2}\,dx-1,
\end{equation}
with $\{y\}=y-\lfloor y\rfloor$, the fractional part of $y\in\R$.
\end{prop}
\begin{proof}
As in the proof of Proposition~\ref{PoisTransf},   ${\cal N}=(N([0,x]))=(t_n)$ denotes a
Poisson point process with intensity $1$. 
By using Equation~\eqref{serie} and by decomposing according to the number of $t_n$'s in
the interval $[0,x]$, one gets 
\begin{align*}
r(x) &= 1+2\sum_{k\geq 0} 2^{k} \P\left(t_2\leq x/2^k\right)\\
&=1+2\sum_{k\geq 0}  2^{k} \sum_{n\geq 2} \P\left(t_2\leq x/2^k, N([0,x])=n\right)\\
&=1+2\sum_{k\geq 0}  2^{k} \sum_{n\geq 2} \P\left(t_2\leq x/2^k\mid N([0,x])=n\right)\Pois{x}{n}.
\end{align*}
According to Proposition~\ref{PoisUnif},  conditionally on the event  $\{N([0,x])=n\}$,
the variable $(t_1, t_2, \ldots t_n)$ has the same distribution as $(xU_{(1)}^n$,
$xU_{(2)}^n$, \ldots, $xU_{(n)}^n)$, where $(U_i, 1\leq i\leq n)$ are independent random
variables uniformly distributed on $[0,1]$ and $(U_{(i)}^n, 1\leq i\leq n)$ denotes their
non-decreasing reordering. In particular, conditionally on the event  $\{N([0,x])=n\}$,
the variable $t_2$ has the same distribution as $xU_{(2)}^n$, hence the quantity
\[
 \P\left(t_2\leq x/2^k\mid N([0,x])=n\right)=
 \P\left(xU_{2,n}\leq x/2^k\right)=
 \P\left(U_{2,n}\leq 1/2^k\right)
\]
does not depend on $x$. The Poisson transform can thus be written as 
\begin{equation}\label{Switch}
r(x)=1+2\sum_{k\geq 0}  2^{k} \sum_{n\geq 2} \P\left(U_{2,n}\leq 1/2^k\right)\Pois{x}{n}
\end{equation}
Now, by switching the two sums, one gets
\begin{align*}
r(x) &=\sum_{n\geq 0} \Pois{x}{n}+ 2\sum_{n\geq 2}\left(\sum_{k\geq 0}  2^{k}\,
\P\left(U_{2,n}\leq 1/2^k\right)\right)\Pois{x}{n}\\ 
&= e^{-x}+xe^{-x}+\sum_{n\geq 2}\left(1+2\sum_{k\geq 0}  2^{k}\, \P\left(U_{2,n}\leq
1/2^k\right)\right)\Pois{x}{n}. 
\end{align*}
The last expression is clearly a Poisson transform, by identifying the coefficients,
this yields the following formula, for $n\geq 2$, 
\[
r_{n}=1+2\sum_{k\geq  0}2^k\, \P\left(U_{2,n}\leq \frac{1}{2^k}\right)
= 1+2 \sum_{k\geq 0} 2^k \,\E\left(\ind_{\{U_{2,n}\leq {1}/{2^k}\}}\right).
\]
By Fubini's Theorem, the sum and the expectation can also be switched (the summands are
non-negative), therefore, 
\[
r_n=1+2 \E\left(\sum_{k\geq 0} 2^k\, \ind_{\{k\leq -\log_2(U_{2,n})\}}\right) 
= 1+2\left(\E\left(2^{\lceil -\log_2(U_{2,n}) \rceil} \right)-1\right).
\]
Since, for $x>0$, $\P(U_{2,n}\leq x)=1-(1-x)^n-nx(1-x)^{n-1}$, the density function 
of the variable $U_{2,n}$ is the function $x\to n(n-1)x(1-x)^{n-2}$ on $[0,1]$. 
This gives the relation
\begin{multline*}
r_n= 4n\int_0^1 2^{\lfloor -\log_2(x) \rfloor} (n-1)x(1-x)^{n-2}\,dx-1\\
= 4n\int_0^1 2^{-\{-\log_2(x)\}} (n-1)(1-x)^{n-2}\,dx-1.
\end{multline*}
The proposition is proved.
\end{proof}

\begin{center}
\begin{figure}[ht]
\resizebox{8cm}{5cm}{\includegraphics{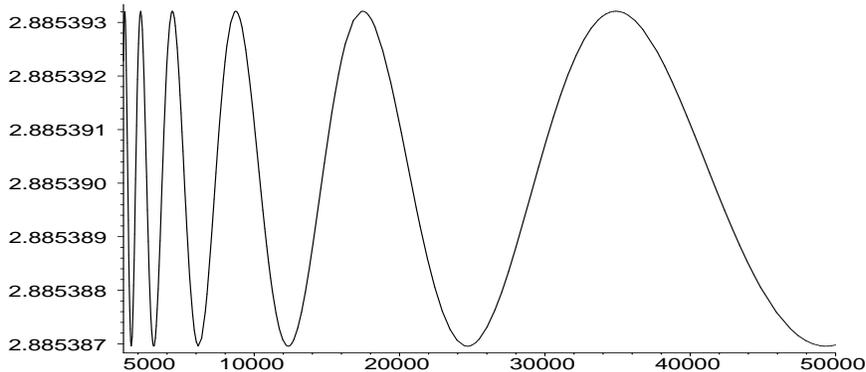}}
\caption{The interpolated sequence $n\to\E(R_n)/n$}
\label{FPeriod}
\end{figure}
\end{center}

With a similar use of Fubini's Theorem as in the above proof, one gets the following
representation of the Poisson transform of $(R_n)$.
\begin{prop}
The Poisson transform  of the sequence $(R_n)$ can be represented as
\begin{equation}\label{Joblot}
r(x) =4x\int_0^{x}2^{-\{\log_2(x)-\log_2(y)\}}e^{-y}\,dy+2(1+x)\,e^{-x} -1, \qquad x\geq 0.
\end{equation}
\end{prop}
\noindent
If Equation~\eqref{Joblot} is more explicit than Equation~\eqref{serie}, it has
nevertheless no use in the asymptotic analysis, Equation~\eqref{serie} is the key identity.  
\begin{proof}
By Equation~\eqref{serie} and Fubini's Theorem, one gets
\begin{align*}
\frac{1}{2}(r(x)-1)&= \sum_{k\geq 0} 2^{k}\,\E\left(\ind_{\{t_2\leq x/2^k\}}\right)
= \E\left(\sum_{k\geq 0} 2^{k}\,\ind_{\{t_2\leq x/2^k\}}\right)\\
&=\E\left[\left(2^{\lfloor\log_2(x/t_2)\rfloor +1}-1\right)\ind_{\{t_2\leq x\}} \right],
\end{align*}
the proof is concluded by trite calculations and by using the fact that the random variable $t_2$ has the density
$(x\exp(-x))$ on $\R_+$. 
\end{proof}
Knuth's result on the asymptotic behavior of this algorithm is given by the following theorem. 
\begin{theo}[Asymptotic oscillations]\label{osc}
The average of $R_n$ satisfies the expansion
\begin{equation}
\E(R_n)=nF(\log_2(n))-1+O\left(ne^{-n}\right), \qquad n\geq 1,
\end{equation}
with 
\begin{equation}
F(y)=4\int_0^{+\infty} 2^{-\{y-\log_2(x)\}} e^{-x}\,dx.
\end{equation}
The function $F$  is periodic with period $1$, in 
particular the sequence $(\E(R_n)/n)$ is not converging.   
\end{theo}
In communication networks, the quantity $\E(R_n)/n$ is related to an average transmission
time when $n$ messages are initially in the network. The practical consequence of the
oscillations of the sequence $(\E(R_n)/n)$ is nevertheless quite limited since the
amplitude of the oscillations of the function $F$ is of the order of $10^{-5}$. See
Figure~\ref{FPeriod}. 
\begin{proof}
By Equation~\eqref{recmoy2}, it is enough to look at the asymptotic behavior of 
\[
\int_0^1 2^{-\left\{-\log_2(x)\right\}}n\left(1-x\right)^{n-1}\,dx=
\int_0^n 2^{-\left\{\log_2(n)-\log_2(x)\right\}}\left(1-\frac{x}{n}\right)^{n-1}\,dx.
\]
The elementary inequality
\begin{multline*}
\left|\int_0^1 2^{-\left\{-\log_2(x)\right\}}n\left(1-x\right)^{n-1}\,dx-
\int_0^{+\infty} 2^{-\left\{\log_2(n)-\log_2(x)\right\}} e^{-x}\,dx\right|\\\leq
\int_n^{+\infty} e^{-x}\,dx+
\int_0^n \left|\left(1-\frac{x}{n}\right)^{n-1}-e^{-x}\right|\,dx=2e^{-n}
\end{multline*}
concludes the proof. 
\end{proof}

\subsection*{Remark}
By evaluating the integral of Equation~\eqref{recmoy2}, one gets that the quantity
$\E(R_n)$ can be expressed as  
\[
\E(R_n) =1+2\sum_{k\geq
  0}2^k\left(1-\left(1-\frac{1}{2^k}\right)^{n}-\frac{n}{2^k}\left(1-\frac{1}{2^k}\right)^{n-1}\right). 
\]
This is the starting point of most of analyses of this algorithm. It is followed by
some exponential approximations, a Mellin transform of the residual series and finally
some complex analysis arguments to derive the asymptotic behavior of the original
sequence. The periodic function $F$ appears then through its Fourier transform. By
inversion, it is expressed in term of the values of the Gamma function $\Gamma$ on a
vertical axis of the complex plane,
\[
F(x)=-\frac{2}{\log 2}\sum_{k\in\Z-\{0\}} \xi_k\Gamma(\xi_k-1)\, e^{2ik\pi x}
\]
with $\xi_k={2ik\pi}/{\log 2}$ for $k\in\Z$. 

Clearly,   it   is   much   easier   to   evaluate  the   asymptotic   behavior   of   the
integral~\eqref{recmoy2}.  Moreover, as a benefit, the periodic function $F$ shows up quite
naturally and with a direct explicit expression which is apparently new. 

\bigskip

\section{Ethernet}\label{Ether}
The context of the algorithm analyzed in this section is the following: In a communication
network with only one  channel, at most one message can be  transmitted each unit of time.
If two transmitters  (or stations) try at the  same time, this is a  failure (a collision)
and both of them have to retransmit later. In an Ethernet network, to each station waiting
for transmission  is associated an integer  $L$ which is  the number of collisions  it has
already  experienced. At  time $t+1$,  a station  with  a counter  equal to  $k$ tries  to
transmit  with  probability  $1/2^{k}$.    See  Metcalf  and  Boggs~\cite{Metcalf:01}  and
Aldous~\cite{Aldous:36}.  For  $t\geq 0$,  $L(t)$ denotes  the value of  the counter  of a
given station which  is waiting for transmission  for $t$ units of time.   In a completely
congested network, the sequence $(L(t))$ evolves as follows: $L(0)=1$ and

\bigskip
\hrule
\[
L(t+1)=
\begin{cases}
L(t) \phantom{ +1 a} \text{ with probability} \qquad 1-a^{L(t)}\\
L(t) +1 \phantom{with probability a} \qquad a^{L(t)},
\end{cases}
\]

\hrule
\bigskip

\noindent
where $a\in(0,1)$. For Ethernet the value of $a$ is $1/2$. 
An approximated counting algorithm proposed by Flajolet and Martin~\cite{Flajolet:17} uses
also such a sequence $(L(t))$, see Flajolet~\cite{Flajolet:16}.

For $k\geq 1$, $G_k$ denotes the sojourn time of $(L(t))$ in state $k$, $G_k$ is a
geometrically distributed with parameter $a^k$:
\[
\P(G_k\geq n) = \left(1-a^k\right)^n.
\]
Consequently,  $G_1+\cdots+G_{k-1}$  is  the  hitting  time  of  $k$  for  the  process
$(L(t))$. Note that the random variables  $(G_k)$ are, of course, independent. 

The average  hitting time of $k$  is therefore $a^{-k}$,  this suggests that the  value of
$L(t)$ is of the order of $\log_{1/a}(t)$ when $t$ is large.  This approximation is indeed
true~: it is not difficult to show that the quantity $|\E(L(t))-\log_{1/a}(t)|$ is bounded
with  respect to $t$.   This estimation  suggests that  there should  be a  convergence in
distribution  of  the  random  variable  $L(t)-\log_{1/a}(t)$ as  $t$  goes  to  infinity.
Surprisingly, this convergence does not  hold.  With  calculus on some alternating series,
Mellin transforms and complex  analysis  methods,  Flajolet~\cite{Flajolet:16}  has shown
that  the  variable $L(t)-\log_{1/a}(t)$ exhibits an asymptotic oscillating  behavior with
respect to $t$. For an extension of these results, see also Kirschenhofer \etal~\cite{Kirschen:01}. 

In this section, a simple proof of  this result is presented, together with  an explicit
description of  the asymptotic behavior of  the distribution of the random variable
$L(t)-\log_{1/a}(t)$. The following elementary proposition simplifies a lot the analysis
of the algorithm. 
\begin{prop}\label{raveneau}
The convergence in distribution 
\begin{equation}\label{Convv}
a^{n}\sum_{k=0}^{n-1} G_{n-k} {\longrightarrow}
H\stackrel{\text{dist.}}{=}\sum_{k=0}^{+\infty} a^k E_k
\end{equation}
holds as $n$ goes to infinity, where $(E_n)$  is a sequence of  independent random variables exponentially distributed
with parameter $1$.
The random variable $H$ has a density $h$ on $\R_+$ given by 
\begin{equation}\label{densh}
h(x)=\frac{1}{\prod_{k=1}^{+\infty}(1-a^k)} \sum_{n=0}^{+\infty} \frac{1}{\prod_{k=1}^{n}(1-a^{-k})}
a^{-n}\exp\left(-a^{-n}x\right), \quad x\geq 0.
\end{equation}
Moreover, Convergence~\eqref{Convv} is true for the total variation norm, i.e.
\begin{equation}\label{Approx}
\lim_{n\to +\infty} \sup_{x\geq 0} 
\left|\P\left(\sum_{k=0}^{n-1} a^nG_{n-k}\geq x\right)-\P\left(H\geq x\right) \right|=0.
\end{equation}
\end{prop}
The random variable $H$ is already known in the domain of communication networks, but in a
very different framework: The congestion avoidance phase of the Transmission Control
Protocol (TCP) of the Internet. In this case, $H$ is related to the stationary
distribution of the throughput of a long TCP connection.  See Dumas \etal~\cite{Dumas:06}. 
The variable $H$ also appeared in mathematical finance models to describe Asian Options,
see Carmona \etal~\cite{Carmona:01}.
\begin{proof}
For $k\geq 0$ and $x>0$, then
\begin{multline*}
\P\left(a^nG_{n-k}\geq x\right)=(1-a^{n-k})^{\lceil x/a^n\rceil}\\=
\exp\left(\rule{0mm}{4mm}\lceil x/a^n\rceil\log\left(1-a^{n-k}\right)\right)\sim \exp\left(\rule{0mm}{4mm}-x/a^k\right),
\end{multline*}
hence the random variable $a^nG_{n-k}$ converges in distribution to $a^k E_k$ where $E_k$
is exponentially distributed with parameter $1$. Therefore, by independence of the
variables $(G_k)$, the convergence~\eqref{Convv} holds. For the explicit expression of the
density of $H$, see the proof of Proposition~13 of Dumas \etal~\cite{Dumas:06}. 

Chebychev's Inequality gives, for $x>0$, 
\[
\P\left(a^{n}\sum_{k=0}^{n-1} G_{n-k}\geq x\right)\leq \frac{1}{x}
\E\left(a^{n}\sum_{k=0}^{n-1} G_{n-k}\right) 
\leq  \frac{1}{x}\frac{a^2}{1-a},
\]
hence, the uniform convergence~\eqref{Approx} has only to be proved in a compact interval.  
Since the sequence $n\to \P\left(a^nG_{n-k}\geq x\right)$ is non-decreasing, the same
property holds for the function $f_n$ defined as follows
\[
n\to f_n(x)\stackrel{\text{def.}}{=}\P\left(a^{n}\sum_{k=0}^{n-1} G_{n-k}\geq x\right).
\]
By Dini's Theorem, the non-decreasing sequence of functions $(f_n)$ converges {\em
  uniformly} on compact sets  to $x\to \P(H\geq x)$. The uniform
convergence~\eqref{Approx} is established .
\end{proof}
\begin{center}
\begin{figure}[ht]
\resizebox{8cm}{5cm}{\includegraphics{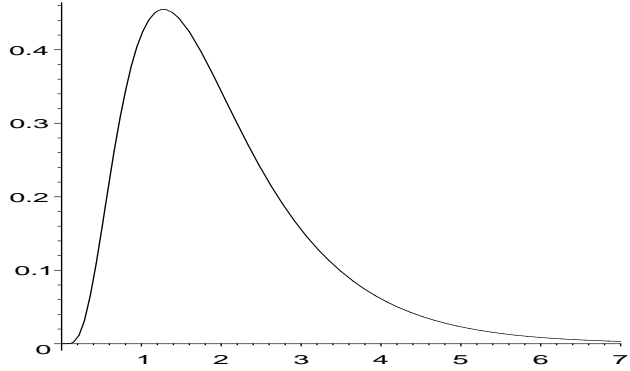}}
\caption{The density function of  $H$ when $a=1/2$}
\label{DensH}
\end{figure}
\end{center}
For $t\geq 1$ and $u\in[0,1]$, it is easy to check that
\[
\E\left(u^{L(t)}\right)=\E\left(\sum_{n\geq 0} u^n(1-u)\, 1_{\{L(t)\leq n \}}\right)
=\sum_{n\geq 0} u^n(1-u)\, \P\left(L(t)\leq n \right)
\]
holds and, since $\P\left(L(t)\leq n\right)=\P\left(G_1+\cdots+G_n\geq t \right)$, for $n\geq 1$.
One thus gets the relation
\[
\E\left[\exp\left(-\lambda L(t)\right)\right]
=\sum_{n\geq 0} e^{-\lambda n}\left(1-e^{-\lambda}\right)\, \P\left(a^{n} \sum_{i=0}^{n-1}
G_{n-i}\geq a^{n} t \right), \quad \lambda>0.
\]
The uniform convergence~\eqref{Approx} gives the expansion, as $t$ goes to infinity, 
\[
\E\left[\exp\left(-\lambda L(t)\right)\right]
=\sum_{n\geq 0} e^{-\lambda n}\left(1-e^{-\lambda}\right)\, \P\left(H\geq a^{n}
t\right)+o(1),
\]
by switching the sum and the expectation, one gets
\begin{align}
\E&[\exp(-\lambda L(t))]\sim \E\left( \sum_{n\geq 0} e^{-\lambda
  n}\left(1-e^{-\lambda}\right)\, 1_{\{H\geq a^{n} t\}}\right)\notag\\
&= \E\left(  \sum_{n\geq \lceil\log_{1{/}a}(t/H)\rceil} e^{-\lambda
  n}\left(1-e^{-\lambda}\right)1_{\{t\geq H\}}\right)\label{EtherFub}\\
&=\E\left[ \exp\left(-\lambda \lceil\log_{1/a}(t/H)\rceil\right) 1_{\{t\geq H\}}\right]
\sim\E\left[ \exp\left(-\lambda \lceil\log_{1/a}(t/H)\rceil\right)\right].\notag
\end{align}
The following equivalence of Laplace transforms has therefore been obtained:
\begin{multline*}
\E\left(\exp\left[-\lambda\left(L(t)-\log_{1/a}(t)\right)\right]\right)
\sim\E\left(\exp\left[-\lambda  \left( \lceil\log_{1/a}(t/H)\rceil -\log_{1/a}(t)\right)\right]\right)\\
=\E\left(\exp\left[-\lambda \left(1-\log_{1/a}(H)-\{\log_{1/a}(t/H)\}\right)\right]\right).
\end{multline*}
The above relation implies that if $x\geq 0$ and  $Z(t)=L(t)-\log_{1/a}(t)$ then
the sequence $(Z(a^{-n-x}))$ converges in distribution. 
The main result of Flajolet~\cite{Flajolet:16} can now be stated.
\begin{theo}[Asymptotic Oscillating Distribution]
Asymptotically, the distribution of $L(t)-\log_{1/a}(t)$ is equivalent to the
distribution of $F(\log_{1/a}(t))$ where, for $x\geq 0$,  $F(x)$ is defined as
\[
F(x)=\log_{1/a}\left(\frac{1}{aH}\right)-\left\{x-\log_{1/a}(H)\right\},\quad x\in\R,
\]
where $H$ is the random variable introduced in Proposition~\ref{raveneau}.
\end{theo}
\noindent
Note that the random function $x\to F(x)$ is periodic with period $1$. 
The variable $H$ is in fact concentrated around its average away from $0$ (see the
detailed study by Litvak and den Zwet~\cite{Litvak:01}) and
from $+\infty$ (exponential decay). The phenomenon of moderate oscillations (but not as
small) seen for Knuth's Algorithm is thus also true in this case. 

\section{Extensions}\label{Gen}

In Section~\ref{Knuth}, the asymptotic behavior of the sequence $(\E(R_n)$ is directly
obtained from Relation~\eqref{recmoy2}. To obtain this identity, the key steps are 1) the
de-Poissonization and 2) the use of Fubini's Theorem in Equation~\eqref{Switch} to remove
the series. In Section~\ref{Ether}, the key step is also the use of Fubini's Theorem in
Equation~\eqref{EtherFub} to get rid of the series. 

It may  be thought that  these derivations are  nevertheless possible only because  of the
particular  expressions involved: For  example, in Section~\ref{Knuth},  the special
properties  of Poisson  processes are  critical  in the  solution of  the problem.   The
purpose of  this section is  to propose a  simple method along the same lines to  study
the asymptotic behavior of some series and functions. As it will be seen, it applies in
various situation, even when there is no probabilistic interpretation of the
sequence/function under study. 

\subsection*{Dyadic Sums}
This example is taken from Flajolet \etal~\cite{Flajolet:14} page~35. The behavior of the function
\begin{equation}\label{Dyad}
G(x)=\sum_{k\geq 1} g\left(x/2^k\right)
\end{equation}
is investigated when $x$ goes to infinity.

It is assumed that $x\to g(x)$ is differentiable and that $g(0)=0$  so that the sum is
well defined. These conditions are not the weakest possible. 
The goal here is to keep the presentation as simple as possible, not to get the most
accurate result for this special case. 

 By using the elementary relation, for $x>0$,
\[
g(x)=\int_0^{x} g'(u)\,du,
\]
the series~\eqref{Dyad} can be written as
\[
G(x)=\sum_{k\geq 1} \int_0^{x} \ind_{\{u<x/2^k\}}g'(u)\,du.
\]
When the sum and the integral are permuted in the last expression, it yields
\[
 \int_0^{x} \sum_{k\geq 1} \ind_{\{u<x/2^k\}} \,g'(u)\,du= \int_0^{x} \lfloor
 \log_2(x/u)\rfloor g'(u)\,du,
\]
Fubini's Theorem states that this last term is indeed $G(x)$ when the condition
\[
\int_0^x |\log_2(u)| \,|g'(u)|\,du <+\infty
\]
holds. In this case, the function $G$ can thus be
represented as
\begin{multline*}
G(x)= \int_0^{x} \lfloor \log_2(x/u)\rfloor \,g'(u)\,du\\=
-\int_0^{x} \left\{\log_2(x/u)\right\} g'(u)\,du+\int_0^{x} \log_2(x/u)\, g'(u)\,du.
\end{multline*}
The following proposition has been proved.
\begin{prop}
If the function $g$ is differentiable and such that $g(0)=0$, 
under the condition
\begin{equation}\label{IntCond}
\int_0^{+\infty} |\log(u)|\,|g'(u)|\,du <+\infty,
\end{equation}
the dyadic sum $G(x)$ can be expressed as
\begin{equation}\label{RepG}
G(x)= F(\log_2(x)) + \int_{x}^{+\infty} \left\{\log_2(x/u)\right\} g'(u)\,du
+\frac{1}{\log 2}\int_{x}^{+\infty} \frac{g(u)}{u}\,du.
\end{equation}
where $F$ is the periodic function, with period $1$,  defined by
\[
F(y)=-\frac{1}{\log 2}\int_0^{+\infty} \frac{g(u)}{u}\,du
-\int_0^{+\infty} \left\{y-\log_2(u)\right\} g'(u)\,du, \quad y\geq 0.
\]
\end{prop}
\noindent
Provided that $g$ has a monotone behavior in the neighborhood of $0$ and $+\infty$,
Condition~\eqref{IntCond} is equivalent to the fact that the integral
\[
\int_0^{+\infty} \frac{|g(u)|}{u}\,du 
\]
exists, i.e. that the Mellin transform of $g$ is defined at $0$. 

From Equation~\eqref{RepG}, it is then not difficult to derive an asymptotic expansion of
$G(x)$ as $x$ goes to infinity.
\subsection*{Harmonic Sums}
More generally, the simple method developed above can be used for a more general class of
functions: 
\[
G(x)=\sum_{k\geq 1} \lambda_k\,g(\mu_k x).
\]
This is the main application  of Flajolet \etal~\cite{Flajolet:14}.  For simplicity, it is
assumed that the sequence $(\mu_k)$ is non-increasing and  converging to $0$  for
example. {\em Provided  that Fubini's Theorem can be applied}, one gets the following
representation: 
\[
G(x)=\int_{0}^{+\infty} \sum_{k\geq 1}  \lambda_k\ind_{\{\mu_k\geq u/x\}}\,g'(u)\, du.
\]
If, for $y>0$ and $n\geq 1$,
\[
\tau(y)=\sup\{k: \mu_k>y\} \text{ and } \Lambda(n)=\sum_{k=1}^n \lambda_k,
\]
with the convention that $\sup\{\emptyset\}=0$, then the harmonic sum can be written as
\[
G(x)= \int_{0}^{+\infty} \Lambda(\tau(x/u))\,g'(u)\, du.
\]
Hence, the rates of growth of $\Lambda$ and $\tau$ give the key of the asymptotic behavior
of the function $G$. 

\subsection*{Final Remark}
It must be noted that, when it can be applied, the method proposed in this paper will require 
weaker assumptions than the  analytic approach.  Indeed,  the conditions to  apply
Fubini's Theorem are  quite minimal (which does not mean that there is no condition at
all).  On the  contrary,  the use  of Mellin  transform implies the  existence of  the
transform itself  on some specified strip of  the complex plane together  with some growth
conditions at infinity. 

\section{Appendix on Poisson Processes}\label{Poissec}
To keep the paper self-contained, this section recalls the basic definitions and results
concerning Poisson processes used in this paper. See  also Kingman~\cite{Kingman:04} 
and Chapter~1 of Robert~\cite{Robert:08} for a more detailed presentation of these
important stochastic processes. 
\begin{defi}
A random variable $X$ has the Poisson distribution with parameter $\lambda$ whenever
\[
\P(X=n)=\Pois{\lambda}{n}, \quad n\in\N.
\]
\end{defi}

The following proposition is the basic property that motivates the use of Poisson
transform when dealing with splitting algorithms. It is a striking elementary property of
the Poisson distribution. 
\begin{prop}[Splitting Property]\label{PoisSplit}
If $X$ is a Poisson random variable with parameter $\lambda$,  $X$ balls are thrown
randomly among $n$ urns and for $1\leq i\leq n$, $X_i$ denotes the number of balls in the
$i$th urn. The variables $(X_i)$  are independent with a common Poisson distribution with
parameter $\lambda/n$.  
\end{prop}

\begin{defi}
A Poisson process with intensity $\lambda$ is an increasing  sequence $(t_n)$  of positive
random variables such that 
\begin{itemize}
\item[---] the increments $t_{n+1}-t_n$, $n\geq 1$,  are independent;
\item[---] For $n\geq 1$ and $x\geq 0$, $\P(t_{n+1}-t_n\geq x)=\exp(-\lambda x)$. 
\end{itemize}
\end{defi}

A Poisson process $(t_n)$ is also represented as a non-decreasing integer valued function
$(N([0,t]), t\geq 0)$ on $\R_+$ where, for $t\geq 0$,  $N([0,t])$ is the number of $t_n$'s 
 in the interval $[0,t]$: $N([0,t])=n$ on the event $\{t_n\leq t<
t_{n+1}\}$. For a Poisson process, the representations as a sequence $(t_n)$ or a
non-decreasing function $(N([0,t]), t\geq 0)$ are  or course equivalent.  

\begin{prop}\label{PoisUnif}
It $(t_n)$ is a Poisson process with intensity $\lambda$, for $t> 0$,
\begin{itemize}
\item[---] the variable $N([0,t])$ has a Poisson distribution with parameter $\lambda$;
\item[---] conditionally on the event $\{N([0,t])=n\}$, the variables 
$$0\leq t_1\leq   t_2\leq\cdots\leq t_{n}\leq t$$ have the same distribution as the
  reordering of  $n$ independent,  uniformly   distributed random variables on $[0,t]$. 
\end{itemize}
\end{prop}

\providecommand{\bysame}{\leavevmode\hbox to3em{\hrulefill}\thinspace}
\providecommand{\MR}{\relax\ifhmode\unskip\space\fi MR }
\providecommand{\MRhref}[2]{%
  \href{http://www.ams.org/mathscinet-getitem?mr=#1}{#2}
}
\providecommand{\href}[2]{#2}

\end{document}